\documentclass[aps,prd,eqsecnum,preprint,tightenlines,nofootinbib]{revtex4-1}
\usepackage{graphicx,latexsym}

\newcommand{\bea}{\begin{eqnarray}}
\newcommand{\eea}{\end{eqnarray}}
\newcommand{\be}{\begin{equation}}
\newcommand{\ee}{\end{equation}}

\newcommand{\ob}[1]{\overline{#1}}
\newcommand{\Pminus}{{\cal P}^-}

\newcommand{\pp}{p^{\prime +}}

\begin{document}

\title{The light-front coupled-cluster method 
applied to $\phi_{1+1}^4$ theory%
\footnote{Based on a talk contributed to the
Lightcone 2014 workshop, Raleigh, North Carolina, 
May 26-30, 2014.}
}


\author{S.S. Chabysheva}
\affiliation{Department of Physics \\
University of Minnesota-Duluth \\
Duluth, Minnesota 55812}

\date{\today}

\begin{abstract}
We use the light-front coupled-cluster (LFCC) method
to compute the odd-parity massive eigenstate of 
$\phi_{1+1}^4$ theory.   A standard Fock-space 
truncation of the eigenstate yields a finite set of linear
equations for a finite number of wave functions.
The LFCC method replaces Fock-space truncation
with a more sophisticated truncation; the eigenvalue
problem is reduced to a finite set of nonlinear equations 
without any restriction on Fock space, but with restrictions
on the Fock wave functions.  We compare our results with 
those obtained with a Fock-space truncation.
\end{abstract}

\maketitle

\section{Introduction} \label{sec:intro}

The nonperturbative solution of quantum field theories in terms of
Fock-state wave functions requires new methods that avoid various
difficulties.  Light-front quantization~\cite{DLCQreviews} is critical 
for this, because it allows for a well-defined Fock-state expansion of
Hamiltonian eigenstates.  The calculation of these wave functions
is usually done in a truncated Fock space, in order to have a finite
number of equations; however, such a truncation brings problems with
uncanceled divergences.  An alternate truncation that apparently
avoids such divergences is made within the light-front coupled-cluster
(LFCC) method~\cite{LFCC}.

The LFCC method replaces a Fock-space truncation
with a more sophisticated truncation, one that limits the way in
which higher Fock-state wave functions are related without completely
eliminating any.  The light-front Hamiltonian eigenvalue problem is 
reduced to a finite set of nonlinear equations, rather than the
finite linear set obtained from a Fock-space truncation.

Here we consider $\phi^4$ theory in 1+1 dimensions as an illustration
of the use of the LFCC method~\cite{LFCCphi4}.  We compute the odd-parity massive 
eigenstate and compare results with those obtained with a Fock-space 
truncation.

Our light-front coordinates~\cite{Dirac} are defined as
$x^+=t+z$ for time and $x^-= t-z$ for space.  The corresponding
light-front energy and momentum are $p^-=E-p_z$ and $p^+= E+p_z$.
The mass-shell condition $p^2=m^2$ becomes $p^-=\frac{m^2}{p^+}$.
The light-front Hamiltonian operator is written as $\Pminus$.

\section{LFCC method} \label{sec:LFCC}

To solve the light-front eigenvalue problem
\be
\Pminus|\psi(P^+)\rangle=\frac{M^2}{P^+}|\psi(P^+)\rangle 
\ee
without making a Fock-space truncation, we
build the eigenstate as 
\be
|\psi\rangle=\sqrt{Z}e^T|\phi\rangle
\ee
from a valence state $|\phi\rangle$ 
and an operator $T$ that increases particle number.
The eigenvalue problem can then be written as
\be
e^{-T}\Pminus e^T|\phi\rangle=e^{-T}\frac{M^2}{P^+}e^T|\phi\rangle.
\ee
We define an effective Hamiltonian $\ob{\Pminus}=e^{-T}\Pminus e^T$,
and the eigenvalue problem becomes
$\ob{\Pminus}|\phi\rangle=\frac{M^2}{P^+}|\phi\rangle$,
which we project onto the valence and orthogonal sectors
\be
P_v\ob{\Pminus}|\phi\rangle=\frac{M^2+P_\perp^2}{P^+}|\phi\rangle, \;\;
(1-P_v)\ob{\Pminus}|\phi\rangle=0.
\ee
with $P_v$ the projection operator.
The second (auxiliary) equation determines $T$.

This formulation is exact; however, in general, $T$ contains an infinite 
number of terms, and the auxiliary equation is really an infinite set
of equations.  The approximation made is to truncate $T$ and truncate
$1-P_v$.  The effective Hamiltonian can then be constructed from a Baker--Hausdorff 
expansion $\ob{\Pminus}=\Pminus+[\Pminus,T]+\frac12 [[\Pminus,T],T]+\ldots $,
which can be terminated when the increase in particle number matches
the truncation of the projection $1-P_v$.

\section{Application to $\phi^4$ theory}  \label{sec:phi4}

The Lagrangian for two-dimensional $\phi^4$ theory is
\be
{\cal L}=\frac12(\partial_\mu\phi)^2-\frac12\mu^2\phi^2-\frac{\lambda}{4!}\phi^4,
\ee
where $\mu$ is the mass of the boson and $\lambda$ is the coupling constant.
The light-front Hamiltonian density is
\be
{\cal H}=\frac12 \mu^2 \phi^2+\frac{\lambda}{4!}\phi^4.
\ee
The mode expansion for the field at zero light-front time is
\be \label{eq:mode}
\phi=\int \frac{dp^+}{\sqrt{4\pi p^+}}
   \left\{ a(p^+)e^{-ip^+x^-/2} + a^\dagger(p^+)e^{ip^+x^-/2}\right\},
\ee
with the modes quantized such that 
\be
[a(p^+),a^\dagger(\pp)]=\delta(p^+-\pp).
\ee

The light-front Hamiltonian is 
$\Pminus=\Pminus_{11}+\Pminus_{13}+\Pminus_{31}+\Pminus_{22}$,
with
\bea \label{eq:Pminus11}
\Pminus_{11}&=&\int dp^+ \frac{\mu^2}{p^+} a^\dagger(p^+)a(p^+),  \\
\label{eq:Pminus13}
\Pminus_{13}&=&\frac{\lambda}{6}\int \frac{dp_1^+dp_2^+dp_3^+}
                              {4\pi \sqrt{p_1^+p_2^+p_3^+(p_1^++p_2^++p_3^+)}} 
     a^\dagger(p_1^++p_2^++p_3^+)a(p_1^+)a(p_2^+)a(p_3^+), \\
\label{eq:Pminus31}
\Pminus_{31}&=&\frac{\lambda}{6}\int \frac{dp_1^+dp_2^+dp_3^+}
                              {4\pi \sqrt{p_1^+p_2^+p_3^+(p_1^++p_2^++p_3^+)}} 
      a^\dagger(p_1^+)a^\dagger(p_2^+)a^\dagger(p_3^+)a(p_1^++p_2^++p_3^+), \\
\label{eq:Pminus22}
\Pminus_{22}&=&\frac{\lambda}{4}\int\frac{dp_1^+ dp_2^+}{4\pi\sqrt{p_1^+p_2^+}}
       \int\frac{dp_1^{\prime +}dp_2^{\prime +}}{\sqrt{p_1^{\prime +} p_2^{\prime +}}} 
       \delta(p_1^+ + p_2^+-p_1^{\prime +}-p_2^{\prime +}) \\
 && \rule{2in}{0mm} \times a^\dagger(p_1^+) a^\dagger(p_2^+) a(p_1^{\prime +}) a(p_2^{\prime +}) .
   \nonumber
\eea
The subscripts indicate the number of creation and annihilation operators
in each term.  Each term changes the number of particles by two or zero, which
allows the eigenstates to be classified as either odd or even in the number
of constituents.

For simplicity of the illustration, we consider the odd case.
The valence state $|\phi\rangle$ is the one-particle state $a^\dagger(P^+)|0\rangle$.
The leading contribution to the $T$ operator is
\be
T_2=\int dp_1^+ dp_2^+ dp_3^+ t_2(p_1^+,p_2^+,p_3^+) 
    a^\dagger(p_1^+)a^\dagger(p_2^+)a^\dagger(p_3^+)a(p_1^++p_2^++p_3^+);
\ee
the function $t_2$ is symmetric in its arguments.  For $T$ truncated to $T_2$, the
projection $1-P_v$ is truncated to projection onto the three-particle state
$a^\dagger(p_1^+)a^\dagger(p_2^+)a^\dagger(p_3^+)|0\rangle$. 
 
Given this truncation, the Baker--Hausdorff expansion for $\ob{\Pminus}$
generates many terms that do not actually contribute to the
valence equation or to the auxiliary equation.  A more efficient approach
for the construction of these equations is to compute only those matrix elements
of $\ob{\Pminus}$ that enter into the projections.
The valence and auxiliary equations become
\be \label{eq:valenceprojected}
\langle 0|a(Q^+)\left(\Pminus_{11}+\Pminus_{13}T_2\right)a^\dagger(P^+)|0\rangle
               =\frac{M^2}{P^+}\delta(Q^+-P^+).
\ee
and
\be \label{eq:auxprojected}
\langle 0|a(q_1^+)a(q_2^+)a(q_3^+)\left(\rule{0mm}{0.2in}
\Pminus_{31}+(\Pminus_{11}+\Pminus_{22})T_2-T_2\Pminus_{11}
     -T_2\Pminus_{13}T_2+\frac12\Pminus_{13}T_2^2\right)a^\dagger(P^+)|0\rangle=0.
\ee
The valence equation can be reduced to~\cite{LFCCphi4} 
\be \label{eq:LFCCvalence}
1+g\int\frac{dx_1 dx_2}{\sqrt{x_1 x_2 x_3}}\tilde t_2(x_1,x_2,x_3)=M^2/\mu^2,
\ee
where $x_i=p_i^+/P^+$,
$g=\lambda/4\pi\mu^2$ is a dimensionless coupling constant,
and $\tilde t_2$ is a rescaled function of longitudinal momentum fractions,
\be \label{eq:tildet2}
\tilde t_2(x_1,x_2,x_3)=P^+t_2(x_1P^+,x_2P^+,x_3P^+).
\ee
We also define a dimensionless mass shift $\Delta$
\be \label{eq:Delta}
\Delta\equiv g\int\frac{dx_1 dx_2}{\sqrt{x_1 x_2 x_3}}\tilde t_2(x_1,x_2,x_3),
\ee
such that $M^2=(1+\Delta)\mu^2$.
The reduced auxiliary equation is~\cite{LFCCphi4}
\bea \label{eq:LFCCaux}
\lefteqn{\frac16\frac{g}{\sqrt{y_1 y_2 y_3}}
             +\frac{M^2}{\mu^2}\left(\frac{1}{y_1}+\frac{1}{y_2}+\frac{1}{y_3}-1\right)
                                    \tilde t_2(y_1,y_2,y_3)}&& \\
   &&  +\frac{g}{2}\left[\int_0^{1-y_1}dx_1
             \frac{\tilde t_2(y_1,x_1,1-y_1-x_1)}{\sqrt{x_1 y_2 y_3 (1-y_1-x_1)}} 
               + (y_1 \leftrightarrow y_2) + (y_1 \leftrightarrow y_3)\right] \nonumber \\
    &&  -\frac{\Delta}{2} \left(\frac{1}{y_1}+\frac{1}{y_2}+\frac{1}{y_3}\right) \tilde t_2(y_1,y_2,y_3)
              \nonumber \\
    && +\frac{3g}{2}\left\{\int_{y_1/(1-y_2)}^1 d\alpha_1 \int_0^{1-\alpha_1} d\alpha_2 
       \frac{\tilde t_2(y_1/\alpha_1,y_2,1-y_1/\alpha_1-y_2) \tilde t_2(\alpha_1,\alpha_2,\alpha_3)}
       {\sqrt{\alpha_1 \alpha_2 \alpha_3 y_3 (\alpha_1-y_1-\alpha_1 y_2)}}\right.      \nonumber \\
    && \rule{2.2in}{0mm} \left.  +(y_1\leftrightarrow y_2)+(y_1\leftrightarrow y_3)
                                               \rule{0mm}{0.15in} \right\} \nonumber \\
    && +\frac{3g}{2}\left\{ \left[ \int_{y_1+y_2}^1 d\alpha_1 \int_0^{1-\alpha_1} d\alpha_2
        \frac{\tilde t_2(y_1/\alpha_1,y_2/\alpha_1,1-(y_1+y_2)/\alpha_1)\tilde t_2(\alpha_1,\alpha_2,\alpha_3)}
        {\alpha_1 \sqrt{\alpha_2 \alpha_3 y_3 (\alpha_1-y_1-y_2)}} \right.  \right.   \nonumber \\
    && \rule{2.5in}{0mm} \left.  + (y_2 \leftrightarrow y_3)
                                                   \rule{0mm}{0.15in}\right]  \nonumber \\
     && \rule{1in}{0mm}  \left. +(y_1\leftrightarrow y_2)+(y_1\leftrightarrow y_3)
                                        \rule{0mm}{0.15in}\right\}=0, \nonumber
\eea
with $y_i=q_i^+/P^+$. 

For comparison, we consider a Fock-state truncation that produces the same 
number of equations.  The truncated eigenstate
\be
|\psi(P^+)\rangle=\psi_1 a^\dagger(P^+)|0\rangle
    +P^+\int dx_1 dx_2 \psi_3(x_1,x_2,x_3)
         a^\dagger(x_1P^+)a^\dagger(x_2P^+)a^\dagger(x_3P^+)|0\rangle
\ee
then contains only one and three-body contributions.
Action of the light-front Hamiltonian $\Pminus$ on this state
yields a coupled system of integral equations, with
$\tilde\psi_3\equiv\psi_3/(\sqrt{6}\psi_1)$:
\bea \label{eq:tildepsi1}
\lefteqn{1+g\int\frac{dx_1 dx_2}{\sqrt{x_1 x_2 x_3}}\tilde\psi_3(x_1,x_2,x_3)=M^2/\mu^2,} && \\
\label{eq:tildepsi3}
&& \frac16\frac{g}{\sqrt{y_1y_2y_3}}
   +\left(\frac{1}{y_1}+\frac{1}{y_2}+\frac{1}{y_3}-\frac{M^2}{\mu^2}\right)\tilde\psi_3(y_1,y_2,y_3) \\
     &&  +\frac{g}{2}\left[ \int_0^{1-y_1} dx_1
             \frac{\tilde\psi_3(x_1,y_1,1-y_1-x_1)}{\sqrt{x_1(1-y_1-x_1)y_2 y_3}} 
               + (y_1 \leftrightarrow y_2) + (y_1 \leftrightarrow y_3)\rule{0mm}{0.3in}\right]
                   =0.   \nonumber
\eea

In each case, the first equation, (\ref{eq:LFCCvalence}) or
(\ref{eq:tildepsi1}), is of the same form; it provides for the self-energy
correction of the bare mass to yield the physical mass.  The second equations,
however, differ significantly.  The LFCC auxiliary equation (\ref{eq:LFCCaux})
includes the physical mass in the three-body kinetic energy; the three-body equation 
of the Fock-truncation approach (\ref{eq:tildepsi3}) has only the bare mass and would
require sector-dependent renormalization~\cite{Wilson,hb,Karmanov,SecDep}
to compensate. The fourth LFCC term is the nonperturbative
analog of the wave-function renormalization counterterm.  The last
two terms are partial resummations of higher-order loops.  These terms
do not appear in the Fock-truncation equation because the loops 
have intermediate states that are removed by the truncation.

\section{Numerical methods}  \label{sec:numerical}

Our numerical method relies
on expansions of $\tilde t_2$ and $\tilde\psi_3$ in a basis of fully symmetric
polynomials~\cite{SymPoly}, which will convert the three-body
equations to systems of nonlinear algebraic equations:
\be
\tilde t_2(x_1,x_2,x_3)=\sqrt{x_1 x_2 x_3}\sum_{n,i}^{n=N} a_{ni}P_{ni}(x_1,x_2).
\ee
The $P_{ni}$ are multivariate polynomials of order $n$
in $x_1$ and $x_2$ that are symmetric with respect
to the interchange of $x_1$, $x_2$, and
$x_3\equiv 1-x_1-x_2$.  The index $i$ distinguishes
between linearly independent polynomials of the same
order; for $n\geq6$ there can be two or more.  The
expansion is truncated at a finite order $N$ so that
the resulting algebraic system is finite in size.

The polynomials $P_{ni}$ can be constructed~\cite{SymPoly}
from linear combinations of
$C_{ml}(x_1,x_2)=C_2^m(x_1,x_2) C_3^l(x_1,x_2)$, where 
$2m+3l\leq n$, and $C_2$ and $C_3$ are given by
\be \label{eq:basepolys}
C_2(x_1,x_2)=x_1^2+x_2^2+x_3^2, \;\;
C_3(x_1,x_2)=x_1 x_2 x_3.
\ee
The most convenient linear combinations are those orthonormal
with respect to the norm 
\be
\int_0^1 dx_1 \int_0^{1-x_1} dx_2\, x_1 x_2 x_3 P_{ni}(x_1,x_2) P_{mj}(x_1,x_2) =\delta_{nm}\delta_{ij}.
\ee

With projection onto the chosen basis functions
$\sqrt{y_1 y_2 y_3}P_n^{(i)}(y_1,y_2)$, the matrix representation
of the auxiliary equation (\ref{eq:LFCCaux}) is found to be
\bea
\lefteqn{\sum_{mj} \left[(1+\Delta)A_{ni,mj}
   -3\left(1+\frac12\Delta\right)B_{ni,mj}+\frac32 g C_{ni,mj}\right]a_{mj}}&& \\
&& \rule{0.5in}{0mm} 
+\sum_{mj}\sum_{lk} \left[9g D_{ni,mj,lk}+\frac92 g F_{ni,mj,lk}\right]a_{mj} a_{lk}
+\frac{g}{6} G_{ni}=0,  \nonumber
\eea
with the self-energy $\Delta$ given by
\be
\Delta=g\sum_{ni}G_{ni}a_{ni}.
\ee
The matrices are  
\bea
A_{ni,mj}&\equiv & \int_0^1 dy_1 \int_0^{1-y_1} dy_2\, y_1 y_2 y_3 P_{ni}(y_1,y_2)
   P_{mj}(y_1,y_2)=\delta_{nm}\delta_{ij}, \\
B_{ni,mj} &\equiv & \int_0^1 dy_1 \int_0^{1-y_1} dy_2 \, y_2 y_3 P_{ni}(y_1,y_2) P_{mj}(y_1,y_2), \\
C_{ni,mj} &\equiv &\int_0^1 dy_1 \int_0^{1-y_1} dy_2 \, y_1 P_{ni}(y_1,y_2) 
                       \int_0^{1-y_1} dx_1 P_{mj}(y_1,x_1), \\
D_{ni,mj,lk}&\equiv & \int_0^1 dy_1 \int_0^{1-y_1} dy_2\, y_1 y_2 P_{ni}(y_1,y_2) \\
&& \rule{0.5in}{0mm} \times
\int_{y_1/(1-y_2)}^1 \frac{d\alpha_1}{\alpha_1} \int_0^{1-\alpha_1} d\alpha_2 
       P_{mj}(y_1/\alpha_1,y_2) P_{lk}(\alpha_1,\alpha_2), \nonumber \\
F_{ni,mj,lk}&\equiv & \int_0^1 dy_1 \int_0^{1-y_1} dy_2 \, y_1 y_2 P_{ni}(y_1,y_2) \\
&& \rule{0.5in}{0mm} \times
\int_{y_1+y_2}^1 \frac{d\alpha_1}{\alpha_1^2} \int_0^{1-\alpha_1} d\alpha_2
        P_{mj}(y_1/\alpha_1,y_2/\alpha_1)P_{lk}(\alpha_1,\alpha_2), \nonumber
\eea
\be
G_{ni}\equiv \int_0^1 dy_1 \int_0^{1-y_1} dy_2 P_{ni}(y_1,y_2).
\ee
They are computed most efficiently by Gauss--Legendre quadrature~\cite{LFCCphi4}.
The same approach applies to the three-body equation of the Fock-space
truncation.

We have tested our numerical method against an analytically solvable case, that of
a restricted three-body problem where the two-two scattering interaction
is dropped from (\ref{eq:tildepsi3}), and found very rapid convergence.
Convergence for the LFCC auxiliary equation is not as rapid, but the 
calculation does converge for a wide range of coupling strengths, using no 
more than the 19 polynomials that occur for $N=12$.  Details can be seen 
in \cite{LFCCphi4}.

\section{Results and summary} \label{sec:results}

\begin{figure}
\vspace{0.2in}
\centering
\includegraphics[width=9cm]{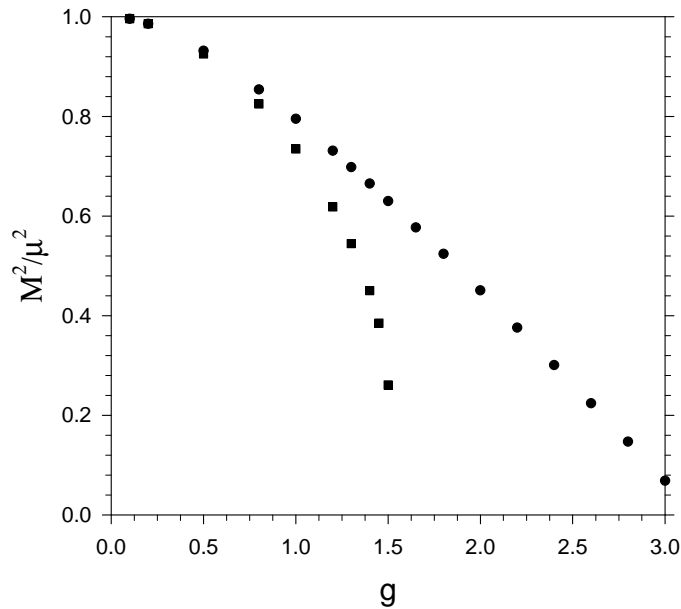}
\caption{Mass-squared ratios $M^2/\mu^2$ versus
dimensionless coupling strength $g$ for the LFCC approximation (squares) and 
the Fock-space truncation (circles).
}
\label{fig:M2vsg} 
\end{figure}

The converged results for the mass-squared eigenvalues are
shown in Fig.~\ref{fig:M2vsg}.  There is a distinct difference
between the LFCC approximation and the Fock-space truncation.
This arises from two factors: the correct kinetic-energy mass
in each sector of the LFCC calculation and contributions from
higher Fock states.
If the Fock-state truncation method is modified with 
sector-dependent masses~\cite{Wilson,hb,Karmanov,SecDep}, 
the resulting mass values are intermediate between the two 
sets shown here~\cite{LFCCphi4}.

To summarize, we have shown an application of the LFCC method to a model theory that 
requires numerical techniques.  Also, suitable techniques have been developed,
based on expansions in fully symmetric polynomials~\cite{SymPoly}.  The results
show important improvements over a Fock-space truncation approach.
This provides a foundation for future work of greater complexity.

Such additional work could include investigation of convergence 
with respect to the terms in the truncated $T$ operator and
analysis of symmetry breaking, for both positive and negative $\mu^2$.
One approach to a study of symmetry breaking would be to consider
the even eigenstates and search for degeneracy of the even and
odd ground states.  At least one additional term in the $T$ operator
would be required, and the even valence state would have two constituents.
A more complete analysis would include zero modes, for which some
preliminary work has already been done~\cite{LFCCzeromodes}.

\acknowledgments
This work was done in collaboration with B. Elliott and J.R. Hiller
and supported in part by the US Department of Energy and the Minnesota Supercomputing Institute.

\end{document}